\documentclass[a4paper, aps, prx,superscriptaddress,twocolumn,a4paper]{revtex4-2}
\usepackage{bm}
\usepackage{graphicx,ulem}

\usepackage{amsmath}
\usepackage{amssymb}%
\usepackage{color}

\usepackage[colorlinks=true,citecolor=blue,urlcolor=blue]{hyperref}
\begin{document}
\title{Approximate universality and large measurement gain of Rabi model in a linear potential under strong Doppler broadening}
\author{Dongyang Yu}
\email{dyyu21@zjut.edu.cn}
\author{Zhan Zheng}

\affiliation{School of Physics, Zhejiang University of Technology, No. 288, Liuhe Road, Hangzhou,310023, Zhejiang, China}

\affiliation{Zhejiang Provincial Key Laboratory and Collaborative Innovation Center for Quantum Precise Measurement, Zhejiang University of Technology, No. 288, Liuhe Road, Hangzhou, 310023, Zhejiang, China}
\begin{abstract}
Harnessing quantum resources in the atomic external degrees of freedom, particularly matter-wave states with large momentum broadening, holds significant potential for enhancing the sensitivity of Kasevich-Chu atom gravimeters at the standard quantum limit. However, a fully quantum-mechanical investigation of the critical Doppler effect inherent to this approach remains lacking. Employing SU(2) Lie group theory, we derive a generic scalar Riccati equation governing the unitary dynamics of the Rabi model within a linear potential and analyze the Doppler effect's impact on Rabi oscillations because of the strong coupling between the internal and external states. Furthermore, by integrating Fisher information theory, we  demonstrate the approximate  universality and high metrological gain of phase-rotation measurement protocols under strong Doppler broadening induced by large-momentum width. This theoretical work provides insightful implications for broader generalization, such as extensions to finite-temperature scenarios or multi-pulse sequences, exemplified by the $\pi/2-\pi-\pi/2$ pulse sequence characteristic of Kasevich-Chu atom gravimeters. Thus this study lays a theoretical foundation for developing high-sensitivity, noise-resistant atom gravimeters that leverage  external-state quantum resources. 
\end{abstract}
\maketitle
\section{Introduction}
The Kasevich-Chu (KC) atom gravimeter~\cite{Kasevich1992,Weiss1994}, based on matter-wave interference, offers significant scientific value and engineering advantages—including high sensitivity and miniaturization~\cite{Berrada2013,Hardman2016,Fu2019,Heine2020}—for applications such as geophysics, mineral exploration, and inertial navigation~\cite{Bidel2018,Zhang2021,Wang2022,Li2023,Chen2025}. 
However, the practical deployment of KC atom gravimeters especially on mobile platforms faces limitations from decoherence mechanisms such as vibration noise~\cite{Merlet2009,Zhang2020}. These not only constrain further sensitivity improvements in traditional KC gravimeters~\cite{Cheinet2008,LeGout2008,Oon2022a,Oon2022b} but also hinder the implementation of many-body entanglement (e.g., spin squeezing) that could substantially enhance the core performance metrics such as  sensitivity~\cite{Hosten2016,Szigeti2020}. 
Consequently, research efforts in recent years have increasingly focused on developing non-entangled quantum resources for atom gravimetry~\cite{Braun2018,Kritsotakis2018}. A prominent example is large-momentum-transfer utilizing Bragg diffraction. Moreover, Recent research~\cite{Kritsotakis2018} indicates that enhancing the sensitivity of atom gravimeters can be achieved not only by improving the phase coherence between the two interferometer paths but also by increasing the momentum broadening of the atomic external state which originates from quantum fluctuations not from thermal broadening, that is, a larger momentum broadening of the input state corresponds to higher sensitivity. 
This research~\cite{Kritsotakis2018} further revealed that, regardless of the chosen measurement scheme, whether population difference, joint momentum-population measurement, or joint position-population measurement, the sensitivity of the conventional KC atom gravimeters retains untapped potential for improvement when supported by optimized parameters and measurement strategies.

Advances in laser cooling technology in recent years have paved the way for realizing KC gravimeters based on a large momentum broadening. It is now feasible to rapidly cool alkali atomic gases to ultra-low temperatures in the sub-microkelvin regime with minimal atom loss~\cite{Chalony2011,Hu2017,YYu2018,Schreck2021,MingjieXin2025}, without requiring evaporative cooling which usually requires a lot of time and produces large atom loss. This capability opens up the possibilities for fully exploiting the quantum resources associated with external atomic states.

Although a large momentum broadening of the input state enhances the sensitivity of KC atom gravimeters, it concomitantly introduces significant Doppler shifts.
Conventionally, Doppler effects are treated semi-classically~\cite{YukunLuo2016,Saywell2018,Fang2018}: quantum fluctuations in the external degrees of freedom are neglected by assuming plane-wave atomic states. The Doppler-induced frequency shift is then incorporated into the two-level atom detuning. The time evolution of the system is subsequently obtained via ensemble averaging using the Maxwell-Boltzmann distribution for thermal atoms. Although this approximation is valid for thermal gases owing to their short de Broglie wavelengths, it breaks down as the temperature decreases. 
Lower temperatures yield longer de Broglie wavelengths, causing quantum fluctuations governed by the position-momentum Heisenberg uncertainty principle to dominate thermal fluctuations. Therefore, the atomic wavefunction manifests as a localized Gaussian wave packet rather than an infinitely extended plane wave.

Understanding the operational characteristics of KC atom gravimeters at elevated momentum broadening  requires resolving  the Doppler-perturbed light-atom interactions. 
Therefore, we investigated the unitary dynamics of the Rabi model in a linear potential~\cite{Lammerzahl1995,Aouachria2002,Benkhelil2023} and its quantum estimation bounds~\cite{Braunstein1994}. For the  complete treatment of external atomic states, particularly the momentum broadening and its corresponding Doppler effects, we employed SU(2) Lie group theory to obtain a generic scalar  Riccati equation~\cite{Markovich2017} essential for optimizing interferometric pulse sequences~\cite{YukunLuo2016}. Additionally, we quantified the Doppler-induced modifications to Rabi oscillations and the relevant quantum Fisher information (QFI).

Concurrently, to fully exploit the quantum resources afforded by large momentum broadening, we introduced a phase-rotation operation (PRO)—implemented by adding a harmonic potential to rotate the atomic momentum and position—for measurement optimization~\cite{Kritsotakis2018}. Using Fisher information theory, we established the (approximate) universality of PRO  measurement and its corresponding performance under both weak and strong Doppler regimes. For the weak Doppler regimes (denoted as "Ideal" scenario through this work), we derived analytical expressions for the classical Fisher information (CFI) corresponding to  joint momentum-population measurements. Our analytical results confirm that optimal measurement can be achieved by tuning the angle of the PRO, saturating the quantum Cram\'er-Rao bound (QCRB), and thus attaining maximum measurement gain~\cite{Braunstein1994}. Crucially, this protocol exhibits universality for any harmonic oscillator eigenstate, meaning that the optimal measurement angle $\theta$ is independent of the quantum number $n$ of the input oscillator eigenstate. This implies that the optimization scheme can be extended to a finite temperature. 

For strong Doppler regimes (denoted as "Doppler" scenario), two critical questions arise naturally: 
Under significant Doppler effects, does the PRO protocol retain the universality? How does the PRO protocol perform? 
To address these issues, we numerically computed the CFI for joint momentum-population measurements. Our numerical results reveal that the universal behavior observed in the "Ideal" scenario remains approximately valid. Furthermore, a substantial measurement gain persists even under strong Doppler effects. 
This study demonstrates the superior robustness of the PRO measurement scheme under strong Doppler effects, thereby providing a solid theoretical basis for future research on anti-decoherence and noise-resistant atom gravimeters, particularly high-sensitivity atom  gravimeters with significant momentum broadening.

This paper is structured as follows. Section II establishes a scalar Riccati equation for arbitrary Doppler strengths and its general unitary evolution operator via SU(2) Lie group theory. Section III investigates the Doppler-strength-dependent evolution of the Rabi oscillations and QFI. Subsequently, Section IV demonstrates the approximate universality of PRO measurement and the substantial enhancement gains in both weak and strong Doppler regimes through Fisher information analysis. Conclusions and outlook complete the study.

\section{SU(2) Unitary dynamics in a linear Potential}
In this section, by employing SU(2) Lie group theory, we derive the generic scalar Riccati equation describing the unitary dynamics of the Rabi model in a linear potential, especially the analytic form of the corresponding unitary operator as the chirping rate of the counter-propagating Raman lasers matches the Doppler shift induced by the linear potential~\cite{Cheng2015}. The Hamiltonian of the  atom-light interaction can be written as follow~\cite{Bord2001}: 
\begin{eqnarray}
	\hat H_0&=&\frac{\hat p^2}{2m}-mg{\hat z}+\hbar\delta(t)|b\rangle\langle b|\nonumber\\
	&+&\frac{\hbar\Omega}{2}e^{i(k_0{\hat z}+\phi)}|b\rangle\langle a|+\frac{\hbar\Omega^*}{2}e^{-i(k_0{\hat z}+\phi)}|a\rangle\langle b|.
\end{eqnarray}
Here, $g$ describes the slope of the linear potential (e.g., the gravitational acceleration constant near Earth's surface). $\delta(t)$ is the two-photon detuning which usually varies with time. $|a\rangle$ and $|b\rangle$ represent the atom’s two internal states, respectively. $\Omega$ is the single-photon Rabi frequency and $\phi$ is the phase difference between the two Raman lasers. $\hbar k_0$ denotes the net momentum transfer induced on the atom during internal state flipping by the two counter-propagating Raman laser beams.

Inspired by Ref.~\cite{XingyuZhang2021}, we sequentially apply two unitary transformations, $\hat U_0=|a\rangle\langle a|\exp{(-ik_0{\hat z}/2)}+|b\rangle\langle b|\exp{(ik_0{\hat z}/2)}$ and $\hat U_1=\exp{(-img{\hat z}t/\hbar)}$, to transition the system into a reference frame where the velocity of atom in state $|a\rangle$ and $|b\rangle$ is $gt\pm\hbar k_0/2$, respectively. In this frame, the momentum becomes a good quantum number, and the Hamiltonian no longer explicitly depends on the position operator $\hat z$. The effective Hamiltonian describing the two-level atom becomes 
\begin{eqnarray}
	{\hat H}_2(t)&=&U_1H_1U_1^\dagger-i\hbar U_1\partial_t U_1^\dagger,\nonumber\\
	&\propto&\frac{({\hat p}+mgt)^2}{2m}+\hbar {\hat B}_3(\hat p,t){\hat S}_z+B_+{\hat S}_+ +B_+^*{\hat S}_-.\nonumber\\
\end{eqnarray}
here $B_+=\Omega e^{-i\phi}/2$ and $\hat B_3(\hat p,t)\equiv -k_0{\hat p}/m-k_0gt-\delta(t)$. 
The primary role of the linear potential is to induce free-fall atomic motion, whereby the momentum increases at a rate proportional to $g$. Consequently, this results in the two-photon detuning to a Doppler shift $k_0gt$. Owing to the implementation of the chirping technique, the two-photon detuning $\delta(t)$ typically exhibits a time dependence. Indeed, the form of $\hat B_3(\hat p,t)$ reveals that the atomic external state also influences the coupling of internal states through the Doppler effect. When the momentum uncertainty $\Delta p$ of the external state satisfies $k_0\Delta p/m\ll |\Omega|$ the external and internal states are nearly decoupled, yielding $B_3(p)\approx-\delta(t)$ ("Ideal" scenario). This regime is applicable to the vast majority of engineering implementations of KC atom gravimeters.
However, enhancement in the sensitivity of KC atom gravimeters may be achieved using quantum states exhibiting high external-state fluctuations or significant momentum broadening, such as SU(1,1) coherent states~\cite{Novaes2004}, characterized by $k_0\Delta p/m\gg |\Omega|$ ("Doppler" scenario). This results in strong coupling between the external and internal atomic states. For scenarios requiring simultaneous consideration of this strong external-internal coupling and chirping modulation, $\hat B_3(\hat p,t)$ generally depends on both the momentum and time. Under these conditions, the unitary dynamics of a two-level atom becomes highly non-trivial. Here, we utilize SU(2) dynamical Lie theory to construct a general Riccati equation governing unitary evolution operator of the system. 

According to SU(2) Lie's group theory, the unitary evolution operator of $\hat H_2$ can be constructed as follows, $\hat U_2= {\tilde U}_2\exp{(-i\int_0^t d\tau (\hat p+mgt)^2/(2m))}$ with 
\begin{eqnarray}
	{\tilde U}_2(t)=\exp{\big(i{\hat f}_+{\hat S}_+\big)}\exp{\big(i{\hat f}_3{\hat S}_3\big)}\exp{\big(i{\hat f}_-{\hat S}_-\big)},
\end{eqnarray}
where $\hat S_\pm$ are the raising/decreasing operator of SU(2) Lie group theory, and $\hat f_\pm(t)$ and $\hat f_3(t)$ depend on the momentum and time, satisfying the following equations:
\begin{eqnarray}
	\frac{d {\hat f}_+(t)}{dt}&=&-B_+-iB_3(t)f_+(t)-B_+^* f_+^2(t),\\
	{\hat f}_3(t)&=&\int_0^t d{\tau} \big(-B_3(\tau)+2if_+(\tau)B_+^*\big),\\
	{\hat f}_-(t)&=&-\int_0^t d{\tau} e^{if_3(\tau)}B_+^*.
\end{eqnarray}
As we can see, the equation of $\hat f_+(t)$ is identical to the scalar Riccati equation~\cite{Markovich2017}, which enables us to investigate the model's evolution at any instant by numerically or analytically computing the unitary operator $\hat U_2(t)$. For instance, they can be utilized to design and optimize composite or shaped pulse schemes~\cite{YukunLuo2016,Saywell2018,Fang2018}. However, this differential equation is a first-order nonlinear equation in time and is dependent on the momentum. Consequently, if the initial state within atom gravimeters possesses large momentum broadening, the development of efficient numerical algorithms becomes particularly crucial. However, under specific parameter regimes, the Riccati equation admits analytical solutions. This is the primary focus of this study.

In KC atom gravimeters, an atomic ensemble undergoes free fall under gravity. The effective coupling between the two internal states $|a\rangle$ and $|b\rangle$ is guaranteed by the  dynamically modulating the two-photon detuning such that $\delta(t)=\delta_0-k_0gt$, that is to say the chirping rate of the counter-propagating Raman lasers  precisely matches the Doppler shift rate induced by the linear potential~\cite{Cheng2015}. Under these conditions, the expression $B_3(\hat p,t)=-\hat B_0$ with $\hat B_0\equiv k_0{\hat p}/m+\delta_0$ becomes explicitly time-independent. Consequently, the unitary operator $\tilde U_2(t)$ corresponding to the Hamiltonian $\hat H_2(t)$ admits an analytical solution:
\begin{eqnarray}
	{\tilde U}_2(t)
	&=&\big({\hat A}(\hat p,t)+i{\hat B}(\hat p,t)\big)|a\rangle\langle a|+\big({\hat A}(\hat p,t)-i{\hat B}(\hat p,t)\big)|b\rangle\langle b|\nonumber\\
	&&-i\hat C({\hat p},t)\big(\exp(i\phi)|a\rangle\langle b|+\exp(-i\phi)|b\rangle\langle a|\big),
\end{eqnarray}
here $\hat A(\hat p,t)=\cos(\hat\Delta t/2)$, $\hat B(\hat p, t)=\hat B_0/{\hat \Delta}\sin(\hat\Delta t/2)$ and $\hat C(\hat p, t)=\Omega/{\hat \Delta}\sin(\hat\Delta t/2)$ with $\hat\Delta=\sqrt{{\hat B}_0^2+\Omega^2}$. In summary, incorporating the Doppler effect, the total unitary operator can be expressed as: 
\begin{eqnarray}
	\hat U(t)=\tilde U_g \hat U_3^{\rm Doppler},\\
	\hat U_3^{\rm Doppler}={\hat U}_0{\tilde U}_2{\hat U}_0^\dagger,
\end{eqnarray}
where $\tilde U_g=\hat U_0 \exp{\Big(-i\frac{t}{\hbar}({\hat p}^2/2m-mg{\hat z})\Big)}\hat U_0^\dagger$. 
For the "Ideal" scenario, $\hat B_0$ becomes negligible. In this limit, $\hat U_3^{\rm Doppler}\rightarrow \hat U_3^{\rm Ideal}$~\cite{Kritsotakis2018}, is given by:
\begin{align}
	{\hat U}_{3}^{\rm ideal}&=\cos(\Omega t/2){\hat I}-i\sin(\Omega t/2)\\
	&\cdot\big(\exp(i(\phi-k_0{\hat z}))|a\rangle\langle b|+\exp(i(k_0{\hat z}-\phi))|b\rangle\langle a|\big).\nonumber
\end{align}
As an initial step in exploring systems with strong external-state quantum fluctuations, we deliberately prepare the initial state as $|\Psi(0)\rangle=|\psi_n\rangle|a\rangle$ and set the single-photon Rabi frequency to $|\Omega|=10E_0$. This simplification facilitates our investigation of the Rabi model dynamics within a linear potential under strong external-internal state coupling. Here, the external state $|\psi_n\rangle$ ($n\in \mathcal{N}$) represents an arbitrary eigenstate of the harmonic oscillator, and $E_0=\hbar^2k_0^2/2m$ denotes the atomic recoil energy. We will conduct a detailed analysis of how momentum broadening and its associated Doppler effects influence Rabi oscillations and the optimization of linear potential gradient measurements.

\section{Doppler-Rabi Oscillations and Quantum Fisher information}
Rabi oscillations constitute one of the most fundamental models in quantum optics and serve as cornerstone technique for manipulating, probing, and utilizing qubits. However, investigations employing a fully quantum methodology to address finite momentum broadening and the concomitant Doppler effect have rarely been reported~\cite{Kozlovskii2001,Banacloche2008}. Beyond flipping the internal atomic states, 
the Raman lasers of the KC atom gravimeter also alter the momentum of the external atomic state. Consequently, momentum broadening in the initial atomic state and the Doppler effect may exert a significant influence on the Rabi model. For example, studies indicate that variations in the external state can cause the final state to deviate substantially from that predicted in the ideal case~\cite{XingyuZhang2021}.

In this section, we employ a fully quantum approach to quantitatively investigate the impact of finite momentum broadening and the associated Doppler effect on Rabi oscillations~\cite{GeaBanacloche2008}. This impact is primarily characterized using two metrics: the probability of $|a\rangle$ state and the final-state fidelity~\cite{XingyuZhang2021}. Specifically, we derive analytical expressions for the difference in QFI between the "Ideal" and "Doppler" scenarios. QFI plays a pivotal role in the subsequent sensitivity analysis.

The probability of $|a\rangle$ state in the final state can be written as,
\begin{eqnarray}
	P_a=\langle\psi_n|{\hat K}_a|\psi_n\rangle,
\end{eqnarray}
here $\hat K_s=\langle a|{\tilde U}^\dagger_2({\hat p}+\hbar k_0/2)|s\rangle \langle s|{\tilde U}_2({\hat p}+\hbar k_0/2)|a\rangle$ for $s=a$ or $b$. 
While the final-state fidelity quantifies the overlap of the wave function at any instant between the “Doppler” and “Ideal” scenarios,
\begin{align}
	{\mathcal F}(t)&=\big|\langle\psi_n|\hat U_3^{\rm Ideal,\dagger}{\hat U}_3^{\rm Doppler}|\psi_n\rangle\big|^2\\
	&=\Big|\langle\psi_n|e^{-i\frac{k_0{\hat z}}{2}}\Big(\cos\left(\frac{\Omega t}{2}\right)\big({\hat A}-i{\hat B}\big)\nonumber\\
	&+\sin\left(\frac{\Omega t}{2}\right){\hat C}\Big)e^{i\frac{k_0{\hat z}}{2}}|\psi_n\rangle\Big|^2.\nonumber
\end{align}
Meanwhile, the analytical form of the QFI, as a key metric for quantifying measurement sensitivity, was calculated. For the “Ideal” scenario, QFI can be simplified as follow: 
\begin{eqnarray}
	F_Q^{\rm Ideal}=4\left(\frac{m^2t^2}{\hbar^2}{\rm Var}({\hat z})+\frac{t^4}{4\hbar^2}{\rm Var}({\hat p})\right).
\end{eqnarray}
While for the “Doppler” scenario $F_Q^{\rm Doppler}$, the QFI difference $\Delta F_Q\equiv F_Q^{\rm Doppler}-F_{Q}^{\rm Ideal}$ can be written in a compacted form, 
\begin{eqnarray}
	\Delta F_Q=4m^2t^2(J_2-J_1^2)+4\frac{mt^3}{\hbar}J_3,
\end{eqnarray}
with
\begin{eqnarray}
	J_1&=&\int dp \big|\langle\psi_n|p\rangle\big|^2(B\partial_p A-A\partial_pB),\nonumber\\
	J_2&=&\int dp  \big|\langle\psi_n|p\rangle\big|^2\big((\partial_p A)^2+(\partial_pB)^2+(\partial_p C)^2\big),\nonumber\\
	J_3&=&\int dp  \big|\langle\psi_n|p\rangle\big|^2 p \big(B\partial_p A-A\partial_pB\big).
\end{eqnarray}

\begin{figure}
	\includegraphics[width=\linewidth]{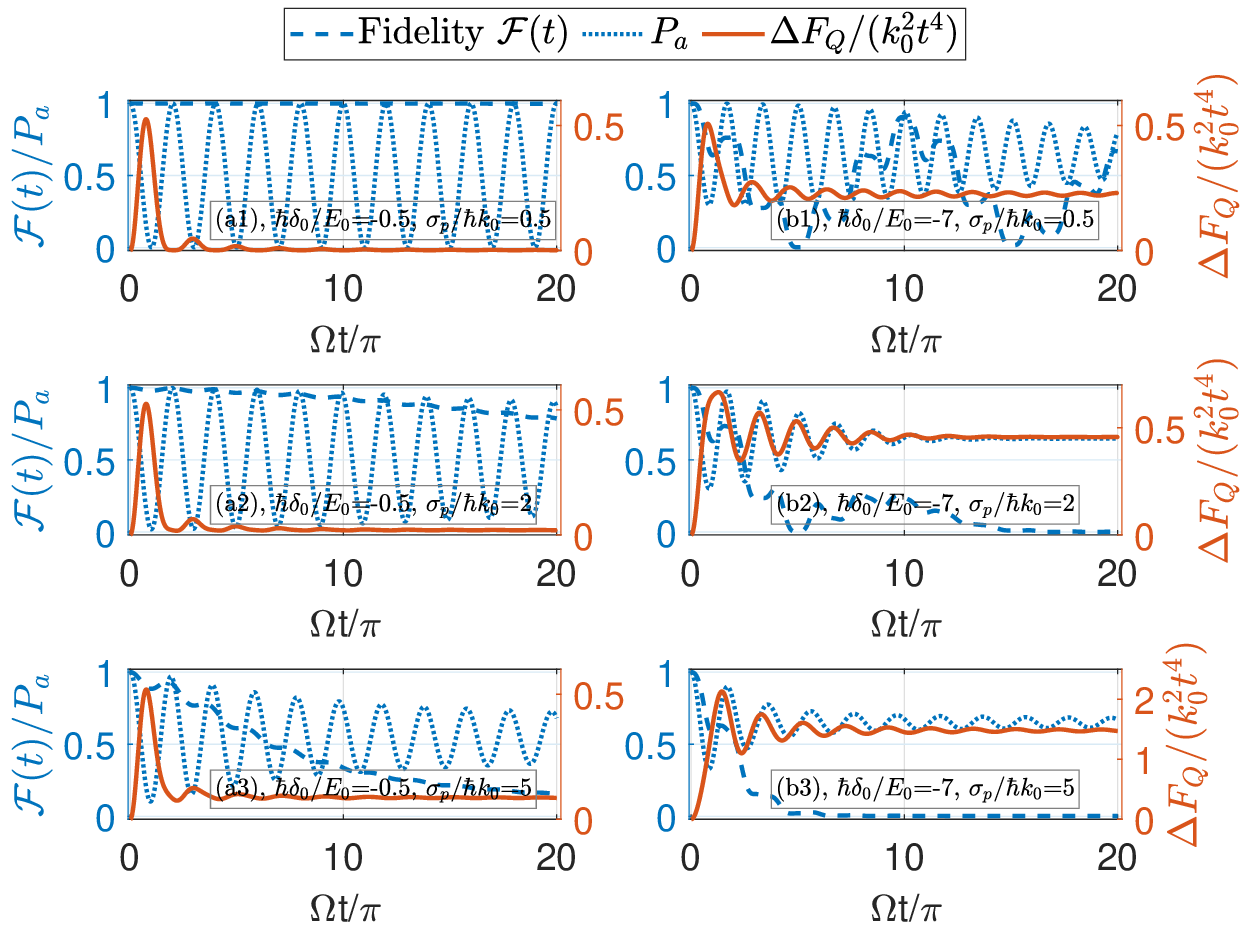}
	\caption{(Color online) The time evolution of the probability of the $|a\rangle$ state $P_a$ (blue dotted lines), the final-state fidelity $\mathcal{F}(t)$ (blue dashed line), and the QFI difference $\Delta F_Q$ (red solid lines) when the ground state of a harmonic oscillator is input, $|\Psi(0)\rangle=|a\rangle|\psi_0\rangle$. The first (second) column corresponds to $\hbar\delta_0=-0.5$ ($-7$)$E_0$, respectively. The momentum broadening varies as $\sigma_p=(0.5, 2, 5)\hbar k_0$ from top to bottom rows. In each panel, the left y-axis corresponds to the probability of the $|a\rangle$ state (blue dotted line) and the final-state fidelity (blue dashed line), while the right y-axis corresponds to the QFI difference (red solid line).}
	\label{fig1}
\end{figure}

Beyond the two-photon detuning $\delta_0$, momentum broadening and the associated Doppler effects significantly influence Rabi oscillations in quantum matter-wave states exhibiting significant momentum broadening, particularly impacting the fidelity of $\pi/2$ or $\pi$ pulses~\cite{Wilkason2022}. To quantitatively illustrate this, we initialize the external state as the harmonic oscillator ground state $|\Psi(0)\rangle=|\psi_0\rangle|a\rangle$ (Fig.~\ref{fig1}). Under two-photon resonance ($\hbar\delta_0=-E_0/2$), the first column of Fig.~\ref{fig1} reveals that the near-perfect Rabi oscillations observed in the narrow momentum-broadening limit deteriorate rapidly as the momentum distribution broadens. Both the amplitude of $\Delta P$ and the final-state fidelity $\mathcal{F}(t)$ diminish substantially with increasing time and a larger value of $\sigma_p$. When the two-photon detuning is tuned away from resonance ($\hbar\delta_0=-7E_0$, second column of Fig.~\ref{fig1}), the amplitude of $\Delta P$ decays at an accelerated rate, whereas the time evolution of the final-state fidelity $\mathcal{F}(t)$ concurrently exhibits pronounced nonlinear behavior. Notably, as shown in Fig.~\ref{fig1}(b), $\mathcal{F}(t)$ can transiently exceed $90\%$ at specific time points. This nonlinear behavior is potentially exploitable for matter-wave beam splitting and recombination in large-momentum-broadening states.

\begin{figure}
	\includegraphics[width=\linewidth]{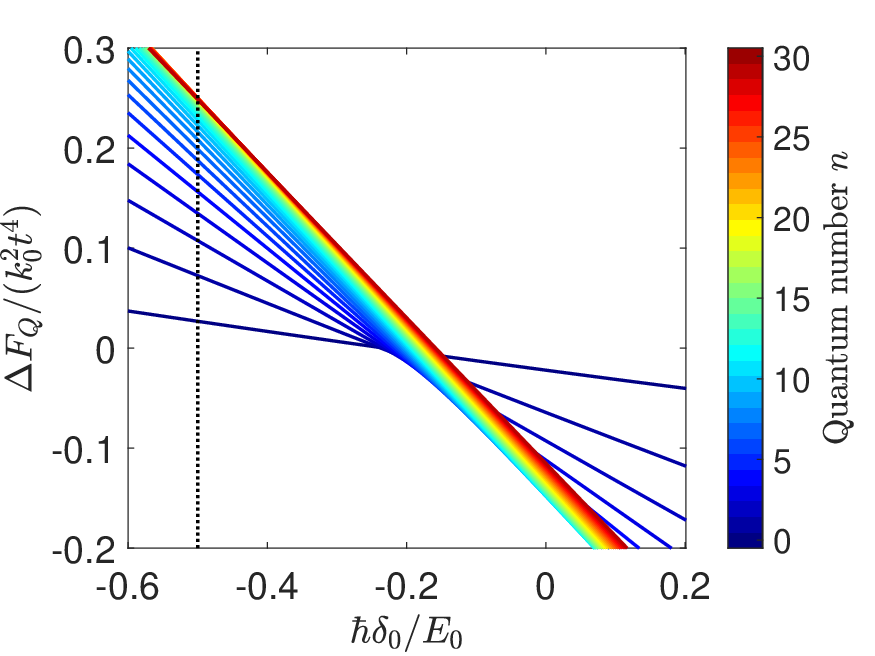}
	\caption{(Color online) In the long-time limit, the slope of QFI difference $\Delta F_Q$ with respect to $\delta_0$ varies as increasing the quantum number $n$ of the eigenstate of a harmonic oscillator. The two-photon resonance is marked by the black dotted line. Here $\Omega t=1000\pi$ and $\sigma_p=2.45\hbar k_0$.}
	\label{fig2}
\end{figure}

For comparison, we also present the temporal evolution of the corresponding QFI difference $\Delta F_Q$ (red solid line in Fig.~\ref{fig1}). Regardless of the values of $\delta_0$ and $\sigma_p$, $\Delta F_Q$ consistently exhibits a characteristic three-stage evolution pattern: an initial phase of rapid growth is followed by a prolonged period of oscillations, which eventually decay, leading $\Delta F_Q$ to asymptotically approach a constant value. This constant is likely to be determined by a combination of quantum fluctuations and $\delta_0$. 
To quantitatively reveal the asymptotic behavior of $\Delta F_Q$ in the long-time limit, we computed its dependence on the two-photon detuning $\delta_0$ and the initial quantum number $n$ at $\Omega t = 1000\pi$, as shown in Fig.~\ref{fig2}. Evidently, $\Delta F_Q$ decreases with increasing the two-photon detuning $\delta_0$, and the decay slope (absolute value) of $\Delta F_Q$ increases with the initial quantum number $n$ until saturation. Notably, under conditions of two-photon resonance, it can be rigorously shown that $J_3=0$.

\section{Classical Fisher Information and Optimized measurements}
According to QCRB theory~\cite{Braunstein1994}, the sensitivity limit for measuring the linear potential slope $g$ is given by the QFI (the value $F_Q$): $\delta g=1/\sqrt{NF_C}\geq 1/\sqrt{NF_Q}$, where $F_C$ represents the CFI corresponding to a specific measurement. In previous studies, researchers either neglected the Doppler effect (the "Ideal" scenario here) to simplify the calculation of Fisher information or treated the Doppler effect semi-classically without comprehensively analyzing its impact on measurement sensitivity from the Fisher information perspective~\cite{Szigeti2012,Kritsotakis2018,Saywell2018}. In this section, we use Fisher information to analyze in detail and demonstrate the approximate universality and corresponding performance of the PRO measurement scheme under strong Doppler effects from momentum broadening. It is worth noting that, due to the momentum-position duality inherent in quantum mechanics, our analysis focuses exclusively on joint momentum-population measurements, omitting the consideration of joint position-population measurements. Specifically, for an initial state corresponding to the harmonic oscillator eigenstate with quantum number $n$, the momentum variance of $|\psi_n\rangle$ is given by $\Delta p=\sqrt{n+1/2}\sigma_p$, where $\sigma_p$ denotes the momentum broadening parameter of the ground state of a harmonic trap.

\subsection{"Ideal" Scenario}
When $k_0\Delta p/m\ll \Omega$, the atomic momentum broadening negligibly affects the atom-Raman light coupling. Consequently, $\hat U_3^{\rm Ideal}$ induces only internal state transitions and momentum transfer, while disregarding Doppler-induced shifts in the effective Rabi frequency. Without PRO implementation, the classical Fisher information $F_{C,p}^{\rm Ideal}$  for joint momentum-population measurement is derived as: 
\begin{eqnarray}
	F_{C,p}^{\rm Ideal}=\left(1+\left(\frac{t\sigma_p^2}{2m\hbar}\right)^2\right)^{-1}F_Q^{\rm Ideal},
\end{eqnarray}
for arbitrary $n$. If the initial momentum broadening $\sigma_p\rightarrow0$, then $F_C^{\rm Ideal}\rightarrow F_Q$. At this limit, the joint momentum-population measurement directly saturates the QCRB. Conversely, $\sigma_p\rightarrow\infty$ or $t\rightarrow\infty$ drives the sensitivity $\delta g\rightarrow\infty$, nullifying the information extraction capability of $g$. However, for the more general case of the momentum broadening parameter $0< \sigma_p< \infty$, it is necessary to optimize the measurement scheme to approach or saturate the sensitivity bound predicted by the QCRB. Because prior research~\cite{Kritsotakis2018} has demonstrated that the PRO scheme can enhance the sensitivity of conventional KC atom gravimeters, where the Doppler effect is negligible under sufficiently low-temperature conditions, we employ Fisher information theory to analyze the universality and performance of the PRO scheme under conditions of strong Doppler effects.
PRO scheme, which is widespread in quantum optics, is divided into two steps: Step 1, Unitary operator $\hat U_s=|a\rangle\langle a|+|b\rangle\langle b|\exp{(-ik_0{\hat z})}$; Step 2,  phase rotation operation: ${\hat U}_{\rm ho}={\hat D}(z_0){\tilde U}_{\rm ho}(\theta){\hat D}^\dagger(z_0)$ and
\begin{eqnarray}
	{\tilde U}_{\rm ho}(\theta)=\exp\left(-i\theta\left({\frac{{\hat p}^2}{2m}}+\frac{m\omega^2}{2}{\hat z}^2\right)\right),
\end{eqnarray}
with translation operator $\hat D(z_0)=\exp{(i{\hat p}z_0/\hbar)}$ and $z_0=\hbar k_0t/(2m)$ diminishing the motion of the center of mass. The resulting CFI for the joint momentum-population measurement, $F_{C,p}^{\rm Ideal}$,
\begin{eqnarray}
	F_{C,p}^{\rm Ideal, HO}
	&=&\frac{t^4}{\hbar^2}\Big(\frac{2}{\omega t\tan(\theta)}-1\Big)^2\int  dp |L_1(p)|^2\Big({\rm Im}\big(\frac{L_2(p)}{L_1(p)}\big)\Big)^2,\nonumber\\
	&=&\frac{(2n+1) \big(m \sigma_p t (t \omega -2 \cot (\theta ))\big)^2}{2 \omega ^2 \left(m^2 \hbar ^2+\sigma_p ^4 t^2\right)+2 \sigma_p ^4 \cot (\theta ) (\cot (\theta )-2 t \omega )}.
\end{eqnarray}
where $L_{1}(p)=\langle p|{\tilde{U}_{\rm ho}}{\hat U}_p|\psi_n\rangle$ and $L_{2}(p)=\langle p|{\tilde{U}_{\rm ho}}{\hat p}{\hat U}_p|\psi_n\rangle$, with $n\in\mathcal{N}$. ${\rm Im}(\cdot)$ indicates taking the imaginary part. 
The analytical result in the second line was verified symbolically for $0\leq n\leq 2$ 
using the \textit{Mathematica} software and numerically for $3\leq n\leq 30$. When tuning the rotation angle $\theta$ to $\theta_{\rm max}^{\rm Ideal}$, $F_Q^{\rm Ideal}=F_{C,p}^{\rm Ideal}$ holds~\cite{ShunLong2006,Facchi2010},
\begin{eqnarray}
	\tan\big(\theta_{\rm max}^{\rm Ideal}\big)=\frac{\sigma_p^4 t}{\omega  \left(2 m^2 \hbar ^2+\sigma_p ^4 t^2\right)},
\end{eqnarray}
Critically, we observe that the optimal rotation angle $\theta_{\rm max}^{\rm ideal}$ is independent of the oscillator quantum number $n$, resulting from the same scaling behavior of $F_{C,p}^{\rm Ideal, HO}$ and $F_Q^{\rm Ideal}$  with respect to $n$, that is $F_{C,p}^{\rm Ideal, HO}\propto 2n+1$ and $F_Q^{\rm Ideal}\propto 2n+1$ or equivalently $F_{C,p}^{\rm Ideal, HO}/F_Q^{\rm Ideal}$=const. for $n\in\mathcal{N}$.  This verifies the universality of the PRO measurement scheme in the "Ideal" scenario, suggesting its robust performance even at finite temperatures.

\subsection{"Dpppler" Scenario}
When $k_0\Delta p/m\geq \Omega$, the Doppler effect from momentum broadening becomes significant, rendering analytical solutions for both QFI and CFI intractable. This raises two critical questions: 1), Universality Persistence, Does the PRO scheme maintain approximate universality, specifically, do both QFI and CFI exhibit similar scaling behaviors with respect to $n$ that closely match those in the "Ideal" scenario? 2), Saturation Robustness, Does the PRO measurement scheme still enable the CFI to saturate its quantum counterpart under strong Doppler effects?

In the absence of the PRO scheme, the CFI for joint momentum-population measurement, $F_{C,p}^{\rm Doppler}$ , can be written as follows:
\begin{align}
	F_{C,p}^{\rm Doppler}=(mt)^2\sum_{s=a}^b\int dp\Big(\frac{\partial_p K_s(p)}{K_s(p)}+\frac{\partial_p P_n(p)}{P_n(p)}\Big)^2
	K_s(p)P_n(p),
\end{align}
here $K_s(p)$ is the eigenvalue in the momentum basis of $\hat K_s$, and $P_n(p)=|\langle p|\psi_n\rangle|^2$. As seen, $F_{C,p}^{\rm Doppler}$ approaches to $F_{C,p}^{\rm Ideal}$ by neglecting the term of $K_s(p)$. 

\begin{figure}
	\includegraphics[width=\linewidth]{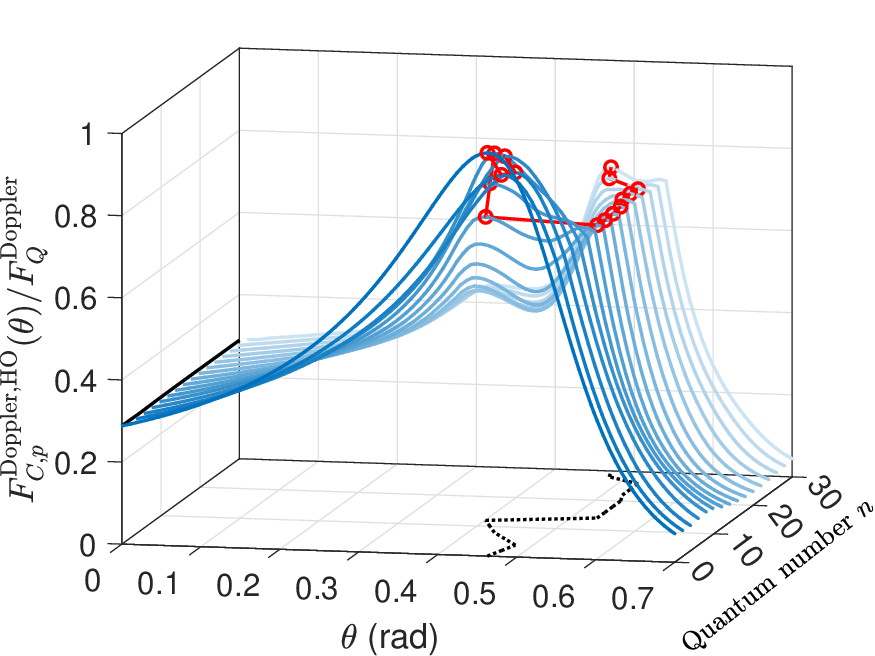}
	\caption{(Color online) The evolution of the CFI for joint momentum-population measurement compared with the QFI, $F_{C,p}^{\rm Doppler, HO}/F_{Q}^{\rm Doppler}$ (red circle symbols, the red solid line serves as a visual guide), as varying the rotation angle $\theta$ and the quantum number $n$. The ratio $F_{C,p}^{\rm Doppler}/F_{Q}^{\rm Doppler}$ without PRO scheme is also plotted as a comparison (black dashed line). Here $\Omega=10E_0$, $\hbar\delta_0=-E_0/2$, $\sigma_p=2\hbar k_0$, $\Omega t=2.5\pi$, and $0\leq n\leq 30$.}
	\label{fig3}
\end{figure}

\begin{figure}
	\includegraphics[width=\linewidth]{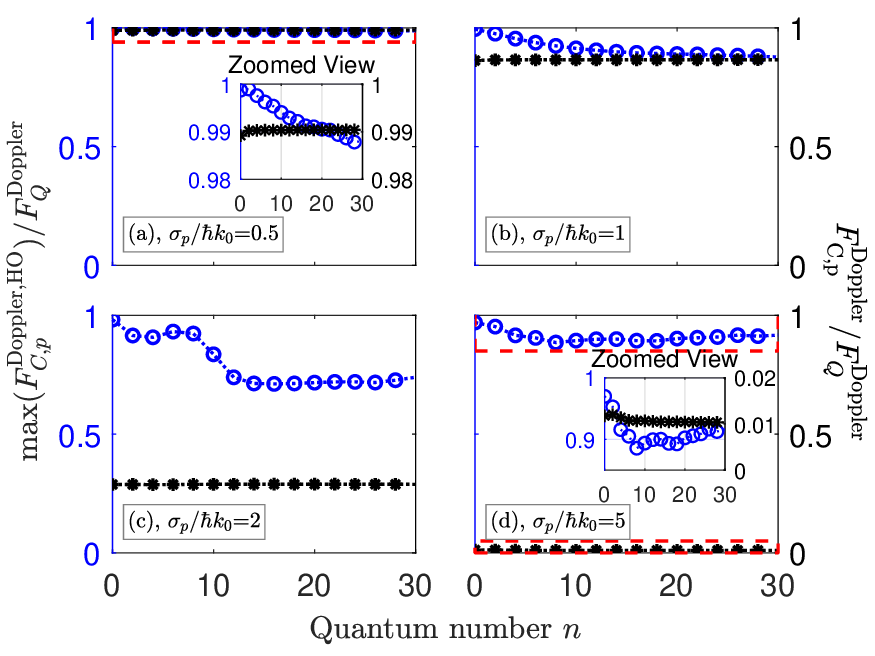}
	\caption{(Color online) The evolution of $F_{C,p}^{\rm Doppler}/F_Q^{\rm Doppler}$ (black asterisk symbols) and  $\max(F_{C,p}^{\rm Doppler, HO})/F_Q^{\rm Doppler}$ (blue circle symbols) as varying the initial momentum broadening parameter $\sigma_p$ and the initial harmonic oscillator quantum number $n$. The dotted lines serve as visual guide. Here $\Omega=10E_0$, $\hbar\delta_0=-E_0/2$ and $\Omega t=2.5\pi$.}
	\label{fig4}
\end{figure}

With PRO implementation, the corresponding CFI can be expressed as follow, 
\begin{align}\label{FcpSz-doppler-HO}
	F_{C,p}^{\rm Doppler, HO}
	&=(\frac{2t}{\hbar\omega\tan(\theta)}-\frac{t^2}{\hbar})^2\nonumber\\
	&\cdot\int dp\sum_{s=a,b}|L_{1,s}(p)|^2\left({\rm Im}\left(\frac{L_{2,s}(p)}{L_{1,s}(p)}\right)\right)^2,
\end{align}
with $L_{1,s}(p)=\langle p|{\tilde U}_{\rm ho}{\hat U}_p\langle s|{\tilde U}_{2}({\hat p}+{\hbar k_0}/2)|a\rangle|\psi_n\rangle$ and $L_{2,s}(p)=\langle p|{\tilde U}_{\rm ho}{\hat p}{\hat U}_p\langle s|{\tilde U}_{2}({\hat p}+{\hbar k_0}/2)|a\rangle|\psi_n\rangle$ for $s=a$ or $b$.
\textit{It is noteworthy that $F_{C,p}^{\rm Doppler, HO}$ involves a double numerical integration, necessitating the use of parallel computing to obtain results efficiently and within a practical timeframe}. We now present the numerical results demonstrating the approximate universality and performance of the PRO scheme under strong Doppler broadening. In Fig.~\ref{fig3}, we set the parameters $\Omega=10E_0$, $\Omega t=2.5\pi$, $\delta_0=-E_0/2$, and $\sigma_p=2\hbar k_0$, followed by scanning the measurement angle $\theta$ and the initial state quantum number $n$ (with $n\leq 30$). Our analysis reveals that, regardless of the specific value of $n$, there always exists an optimal measurement angle $\theta_{\rm max}^{\rm Doppler}$ that drives $F_{C,p}^{\rm Doppler, HO}$ as close as possible to the QFI ($F_Q^{\rm Doppler}$) (even if the QCRB is not saturated). Furthermore, $\theta_{\rm max}^{\rm Doppler}$ oscillates around the mean value as $n$ varies, as shown by the black dotted curve in Fig.~\ref{fig3}. This indicates that the approximate universality of the PRO scheme is largely preserved even under strong Doppler broadening (characterized by $\sigma_p\sqrt{n+1/2}\geq \Omega$). This preservation originates from the fact that both the QFI and CFI under strong Doppler broadening maintain the scaling behaviors with respect to the initial quantum number $n$ which are qualitatively similar to those observed in the ideal scenario (see Fig.~\ref{fig4})~\cite{ShunLong2006,Kritsotakis2018}. Finally, we observe that the PRO scheme delivers substantial measurement gains, often by factors of several times or more, compared to the scenario without PRO optimization, as evidenced by the comparison between the red circular data points and the black dashed line in Fig.~\ref{fig3}. 

To explicitly elucidate the impact of the Doppler broadening strength on the universality and metrological gain of the PRO scheme, Fig.~\ref{fig4} presents the evolution of the ratio of the maximum CFI for the joint momentum-particle number measurement to the QFI, $\max(F_{C,p}^{\rm Doppler, HO})/F_Q^{\rm Doppler}$, as a function of the initial momentum broadening parameter $\sigma_p$ and the initial harmonic oscillator quantum number $n$. For comparison, the corresponding ratio $F_{C,p}^{\rm Doppler}/F_Q^{\rm Doppler}$ without the PRO scheme is also plotted. Notably, this ratio exhibits negligible dependence on the initial quantum number $n$ (indicated by the black asterisk symbols in Fig.~\ref{fig4}). When $\sigma_p=\hbar k_0/2$ (Fig.~\ref{fig4}(a)), where Doppler broadening is almost negligible, the ratio $\max(F_{C,p}^{\rm Doppler, HO})/F_Q^{\rm Doppler}$ for the PRO measurement remains nearly constant with respect to $n$. However, the enhancement provided by the PRO scheme is marginal. As $\sigma_p$ increases, the Doppler effects become progressively more significant. While the CFI corresponding to the PRO scheme begins to exhibit some variation with $n$, the ratios $\max(F_{C,p}^{\rm Doppler, HO})/F_Q^{\rm Doppler}$ maintain their approximate constancy. Concurrently, the metrological gain achieved by the PRO measurement starts to increase (compare the black asterisk symbols and blue circular data points in Fig.~\ref{fig4}). At $\sigma_p=5\hbar k_0$  (Fig.~\ref{fig4}(d)), where the Doppler broadening is strong, the PRO scheme still preserves an approximately constant CFI to QFI ratio across a broad range of $n$ ($0\leq n\leq 30$). Crucially, a substantial enhancement in measurement sensitivity is achieved. In summary, our Fisher information analysis establishes the approximate universality and robust metrological gain of the PRO scheme under strong Doppler broadening. This implies that a PRO scheme delivers significant sensitivity enhancements in thermal atomic gases across the entire Doppler broadening spectrum—from weak to strong regimes~\cite{Szigeti2012}.

\section{Conclusion and Outlook}
We derived the unitary evolution operator for the Rabi model in a linear potential analytically using the SU(2) Lie group theory. This derivation yields a generic scalar Riccati equation governing the unitary dynamics of a two-level atom, which plays a crucial role in designing or optimizing the pulse shapes or sequences for quantum matter-wave states with large momentum broadening. Building upon this theoretical framework, we investigated the Doppler effect of momentum broadening by analyzing its impact on Rabi oscillations and the sensitivity of linear potential slope measurements. Remarkably, we found that even under strong Doppler broadening, the approximate universality of the phase-rotation operation and substantial metrological gain are largely preserved, comparable to weak-Doppler regimes. Although our current analysis focuses on the unitary dynamics of the Rabi model in a linear potential, our methodology readily extends to multi-pulse scenarios. This extension provides a theoretical foundation for harnessing external-state quantum resources exhibiting strong quantum fluctuations~\cite{Robins2013}, such as the SU(1,1) coherent states~\cite{Novaes2004,Lv2020} characterized by ultra-large momentum broadening, within Kasevich-Chu atom gravimeters.

\section{Acknowledgments}
The authors thank Fong en Oon and Chenwei Lv for stimulating discussions and declare that they have no conflicts of interests. 
D.Y. was supported by the start-up fund of the Zhejiang University of Technology. This manuscript has no associated data or the data will not be deposited. [Authors comment: The datasets generated and/or analyzed during the current study are available from the corresponding author on reasonable request.]

\bibliography{mybib}
\end{document}